\documentclass[11pt,twoside]{article}

\usepackage{asp2006}
\usepackage{epsf}
\usepackage{psfig}
\usepackage{lscape}

\begin{document}

\title{Simulations of Supersonic Turbulence in Molecular Clouds: Evidence for a New Universality}   

\author{A. G. Kritsuk\altaffilmark{1}, S. D. Ustyugov\altaffilmark{2}, M. L. Norman\altaffilmark{3}, 
and P. Padoan\altaffilmark{3}}

\altaffiltext{1}{Center for Ap. \& Sp. Sci., UC San Diego,
9500 Gilman Drive, La Jolla, CA 92093-0424, USA;
Department of Math. \& Mech., St. Petersburg State University, St. Petersburg 198504, Russia}
\altaffiltext{2}{Keldysh Institute for Applied Mathematics, Russian Academy of Sciences,
Miusskaya Pl. 4, Moscow 125047, Russia}
\altaffiltext{3}{Physics Department, UC San Diego,
9500 Gilman Drive, La Jolla, CA 92093-0424, USA}

\begin{abstract} 
We use three-dimensional simulations to study the statistics of supersonic 
turbulence in molecular clouds. Our numerical experiments describe driven 
turbulent flows with an isothermal equation of state, Mach numbers around 10, 
and various degrees of magnetization. We first support the so-called 1/3-rule
of \cite{kritsuk...07a} with our new data from a larger $2048^3$ simulation.
We then attempt to extend the 1/3-rule to supersonic MHD turbulence and
get encouraging preliminary results based on a set of $512^3$ simulations.
Our results suggest an interesting new approach to tackle universal scaling 
relations and intermittency in supersonic MHD turbulence.
\end{abstract}

\section{Introduction}
Supersonic turbulence plays an important role in shaping hierarchical internal 
substructure of molecular clouds (MCs). Since the process of star formation
begins with the formation of dense cores, turbulence can be responsible for
creating initial conditions for star formation. It is hard to access the
supersonic regimes typical for MC turbulence in the laboratory and the information
available from observations is also limited. Thus numerical simulations currently
represent the only way to explore the statistics of supersonic turbulence, the
key ingredient of any successful statistical theory of star formation 
\citep{padoan...07,mckee.07}. For a long time, the nature of {\em highly compressible}
and {\em magnetized} interstellar turbulence remained poorly understood. We use
large-scale numerical simulations to shed light on the energy transfer between
scales and on the key spatial correlations of relevant fields in these flows.

\section{The 1/3-rule for Supersonic Hydrodynamic Turbulence}
In \citet[][hereafter K07]{kritsuk...07a} we showed that homogeneous isotropic
turbulence in a strongly compressible regime typical for the interstellar gas
can be approximated by modified Kolmogorov's laws since 
nonlinear advection still remains the major physical process at work in the
absence of magnetic fields. The essence of the required modification boils down
to the use of proper density weighting for the gas velocity, which we call
the ``1/3-rule'' of supersonic turbulence. Replacing the velocity ${\bf u}$ by
$\rho^{1/3}{\bf u}$, one can show that for this quantity the 4/5-law of 
\citet{kilmogorov41} approximately holds at sonic Mach number as large as $M_s=6$, 
when the plain velocity power spectrum already scales as in Burgers turbulence, 
i.e. $\sim k^2$.
\begin{figure}
\centering
\plotone{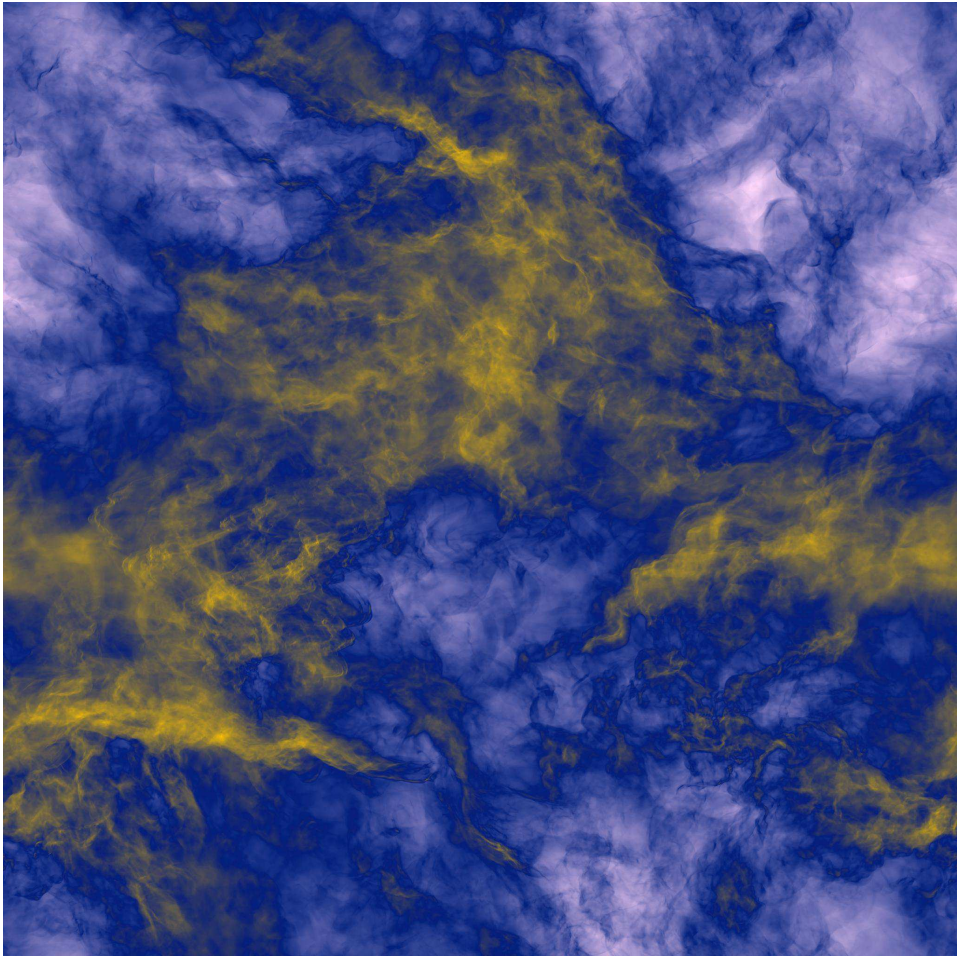}
\caption{\small A snapshot of the projected gas density from the $2048^3$ Mach 6 turbulence 
simulation with PPM. White-blue-yellow colors correspond to low-intermediate-high projected density values.
}
\vspace{-0.5cm}
\label{figone}
\end{figure}
The first von~K\'arm\'an--Howarth (1938) relation between the second-order 
transverse and longitudinal velocity structure functions also still holds 
approximately at $M_s=6$. Although there is no rigorous proof for any of these 
two fundamental exact incompressible laws in supersonic regimes, the evidence 
provided by numerical experiments helps %to 
extend Kolmogorov's phenomenology 
to a wider class of problems of interest in turbulence research and astrophysics.
In K07 we showed that the low-order statistics of $\rho^{1/3}{\bf u}$ are 
invariant with respect to changes in the Mach number. For instance, the slope 
of the power spectrum is~$-1.69$, and the exponent of the third-order 
structure function $S_3(\ell)$ is unity, $S_3(\ell)\equiv\left<|\delta u({\ell}) |^3\right>\sim\ell$.
%, where $\delta u({\ell})$ is the velocity difference between two random positions
%separated by a linear distance $\ell$ and the brackets denote an ensemble average.

Here we report on our more recent simulation with a grid resolution of $2048^3$
carried out with the ENZO code on {\em Bigben} (PSC) and on {\em Ranger} (TACC) using 4096 cores
(Fig.~\ref{figone}). The simulation details can be found in K07, except that this new 
run employs a purely solenoidal large-scale driving force.

\begin{figure}
\centering
\plottwo{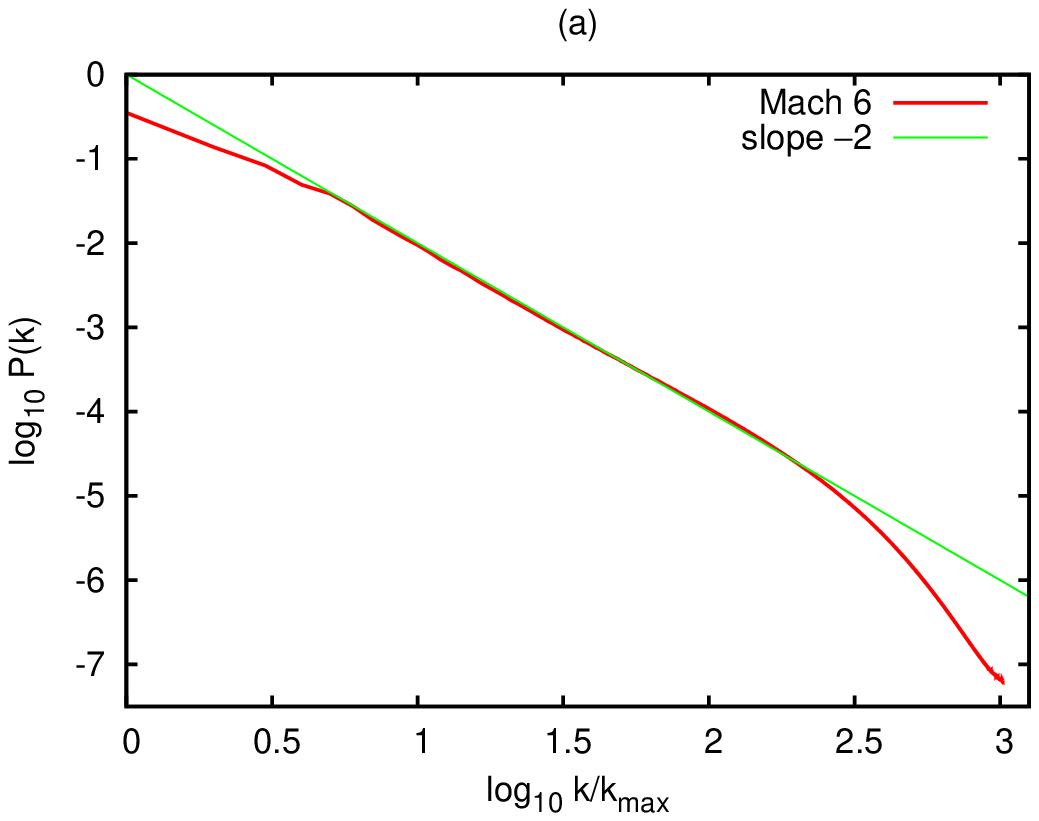}{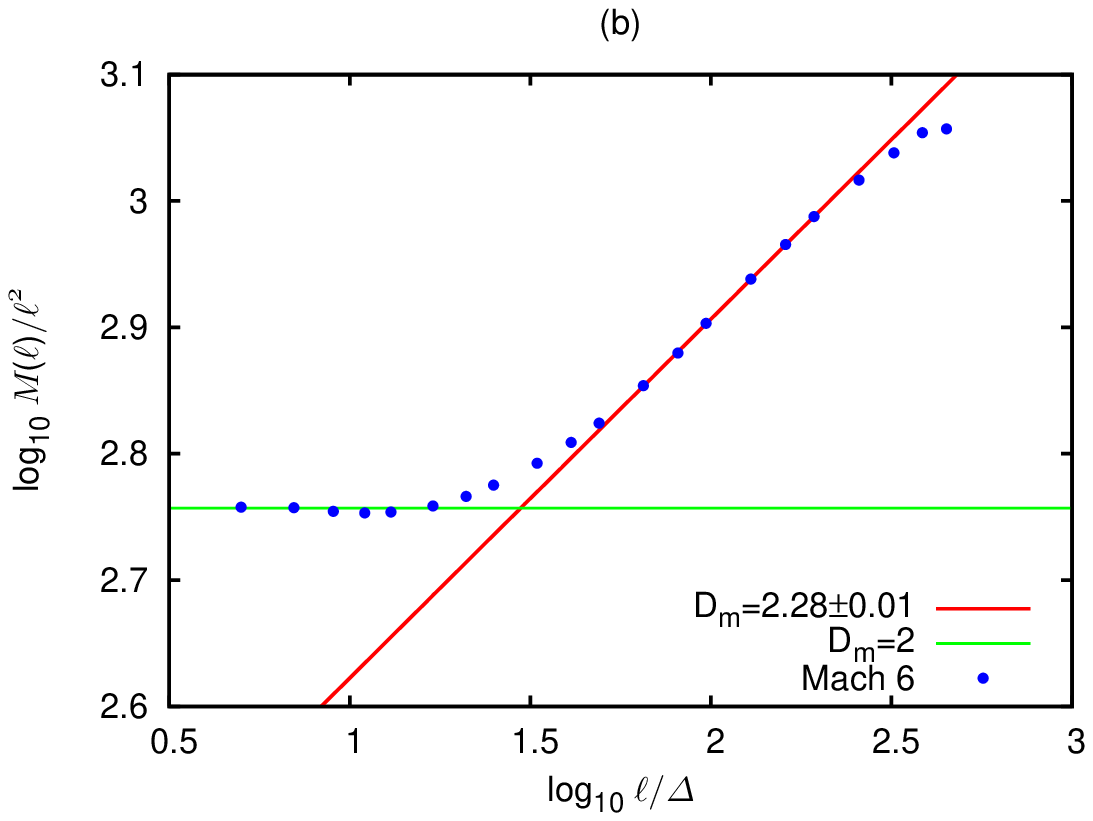}
\plottwo{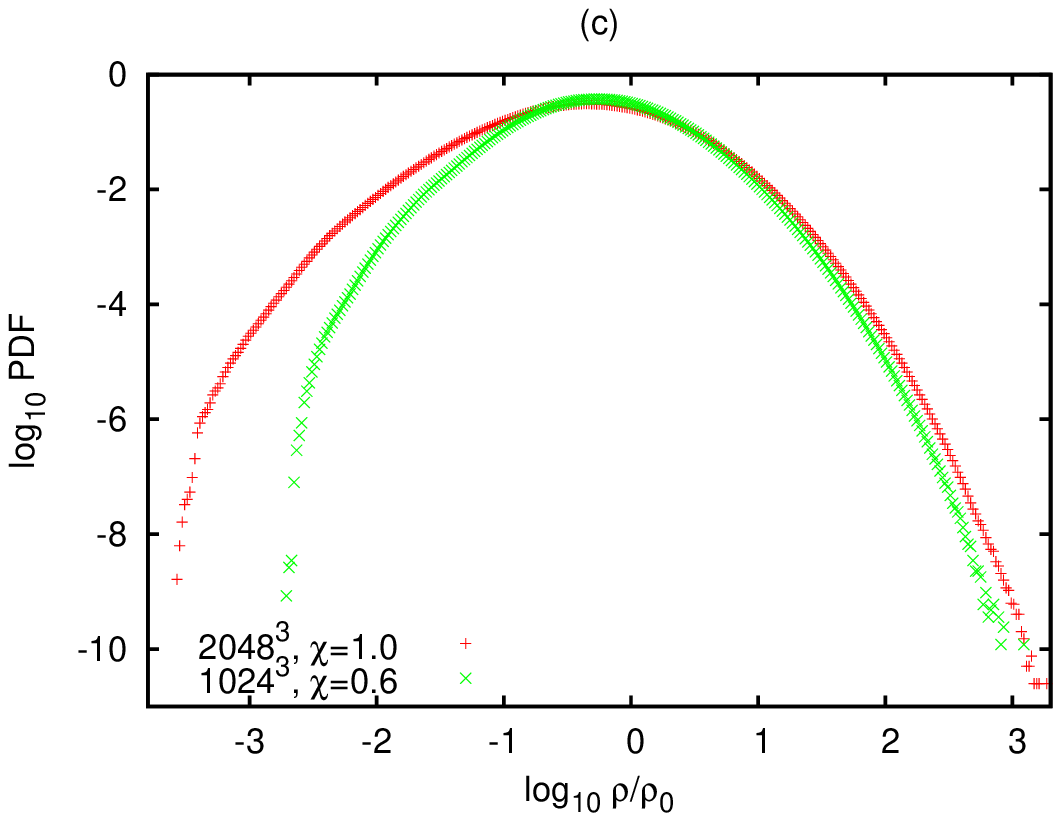}{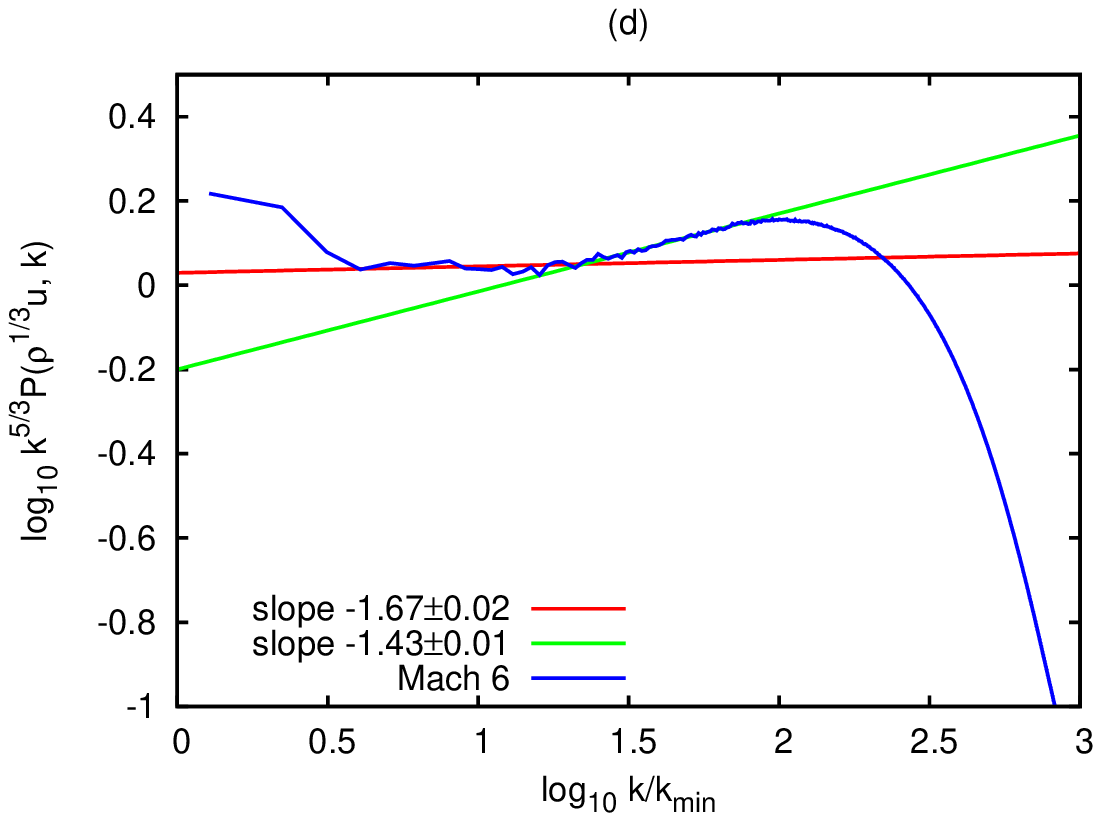}
\caption{\small Statistics of supersonic turbulence from the $2048^3$ simulation at 
$M_s=6$: ({\bf a}) power spectrum of projected density (an ensemble average over 1275 projections);
({\bf b}) average gas mass $M(\ell)$ for boxes centered around the highest density peaks as a 
function of box size, the horizontal line corresponds to a fractal mass dimension $D_m=2$;
({\bf c}) density pdfs for the $2048^3$ simulation with a solenoidal driving force ($\chi=1$)
and for $1024^3$ simulation with a hybrid driving ($\chi=0.6$, K07); 
({\bf d}) compensated power spectrum of $\rho^{1/3}{\bf u}$.
}
\vspace{-0.5cm}
\label{figtwo}
\end{figure}

Turbulence statistics derived from these new data
with unprecedented dynamic range confirm the convergence of our earlier results obtained 
with lower resolution. Figure~\ref{figtwo} provides a brief summary of the new 
statistics. The power spectrum of projected gas density has a slope of $-2.01\pm0.01$
in the inertial range, consistent with the 3D density power index of $-1.07\pm0.01$
reported in K07 (Fig.~\ref{figtwo}{\bf a}). 
The slopes of the power spectra of density and log density (not shown) are very close
to $-1$ and to $-5/3$, respectively. However, it is a mere coincidence that at Mach 6 the 
slopes are represented by these ``good'' numbers, since already at $M_s=10$ they become 
shallower. Our new estimate for the ``fractal'' dimension of the density distribution for 
the inertial range of Mach 6 turbulence, $D_m\approx2.3$, agrees well with the previously 
obtained value of 2.4 (Fig.~\ref{figtwo}{\bf b}, see K07 for definition details).
In the dissipation range, where individual shock fronts are the dominant structures, $D_m=2$
\citep[cf.][]{kritsuk...07b,schmidt..08,pan..09,schmidt....09}. Figure~\ref{figtwo}{\bf c} 
compares the probability density functions (pdfs) of the gas density from simulations 
with different driving forces. Both cases agree nicely with lognormal distributions at 
high densities and show some divergence at low densities that can be at least partly 
attributed to poor statistical representation of violently evolving rarefactions due to a 
limited number of flow snapshots used to derive the pdfs (100 for the $2048^3$ data and 
170 for K07 data), cf. \citet{lemaster.08,federrath..08}. Finally, Fig.~\ref{figtwo}{\bf d}
shows a sample power spectrum of $\rho^{1/3}{\bf u}$ that also confirms the scaling
obtained in K07 based on a lower resolution simulations.

\section{New Scaling Laws for Supersonic MHD Turbulence}
\begin{figure}
\centering
\plottwo{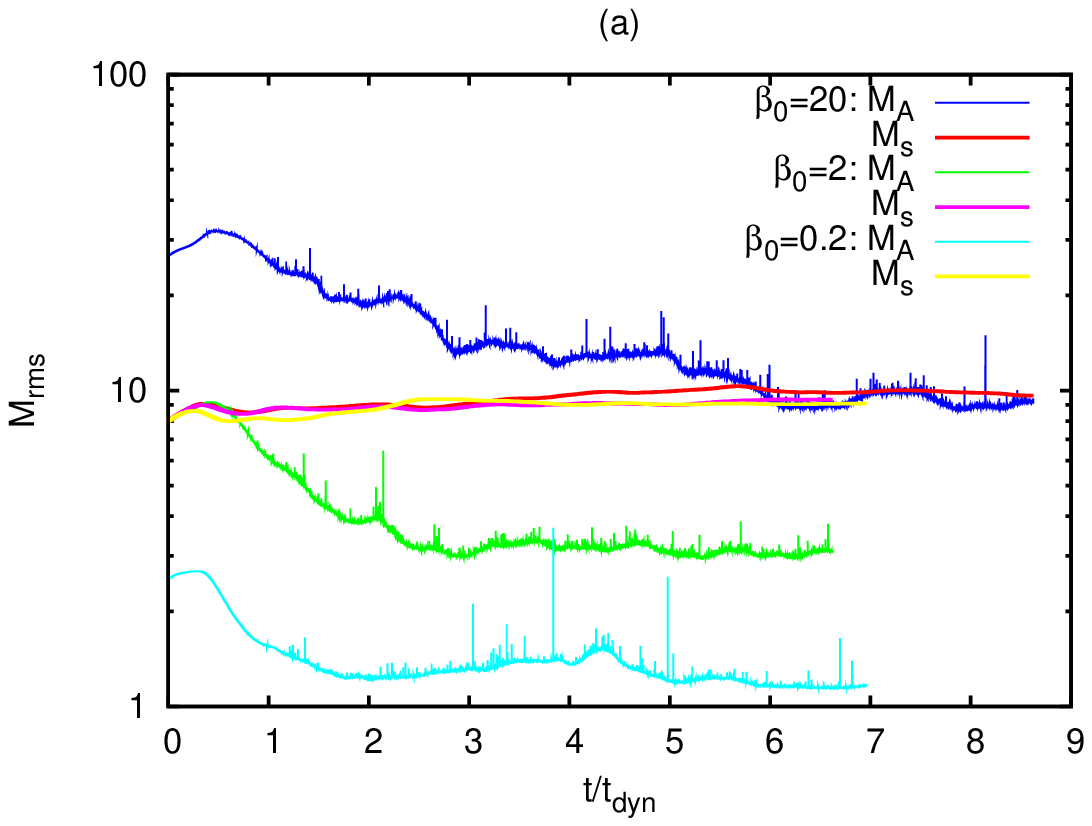}{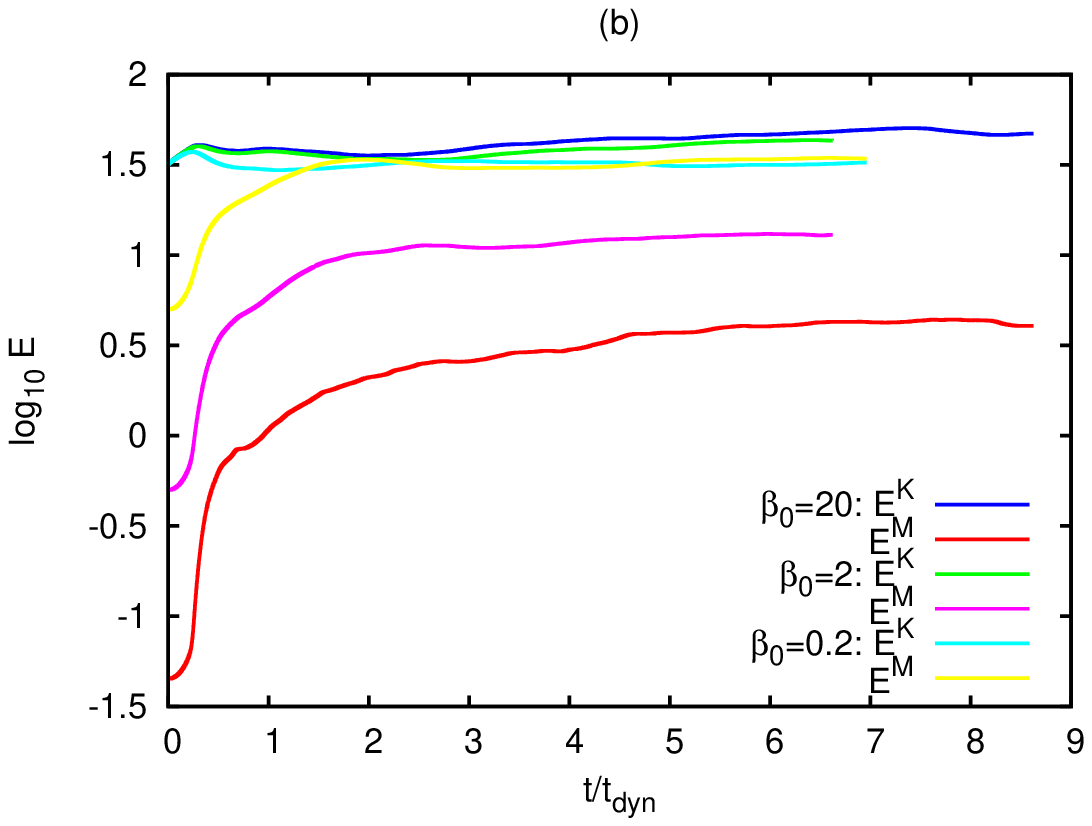}
\plottwo{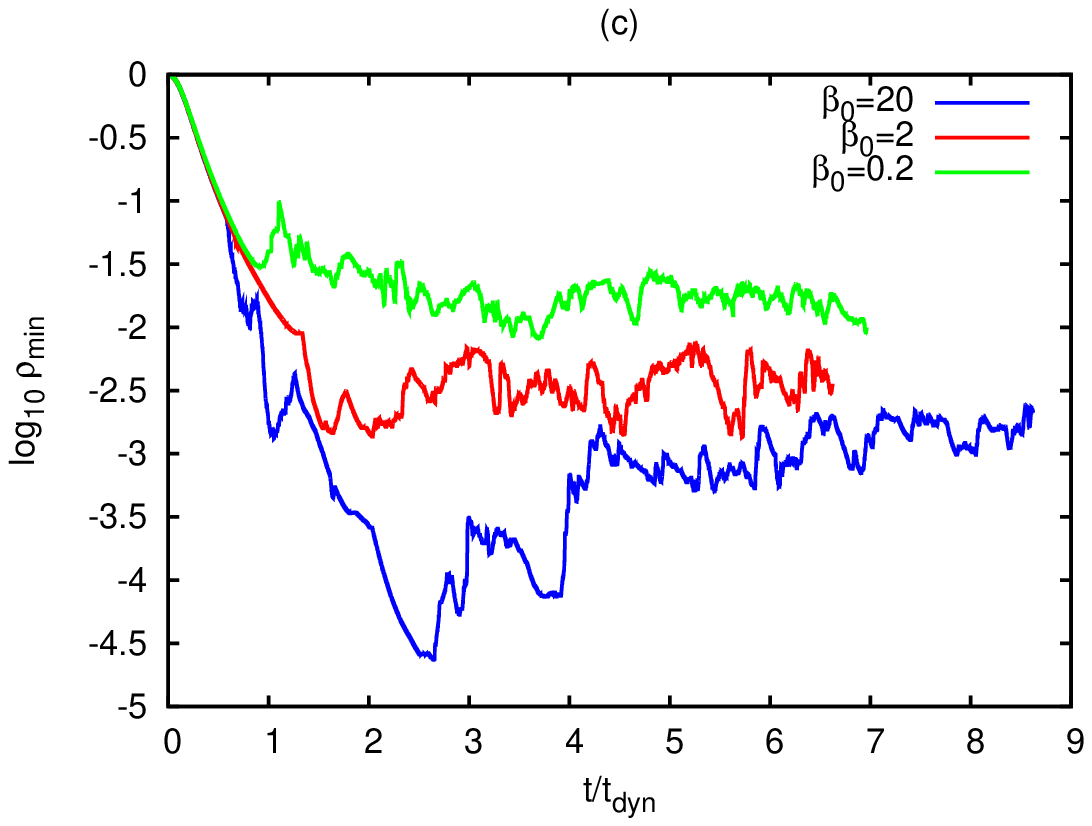}{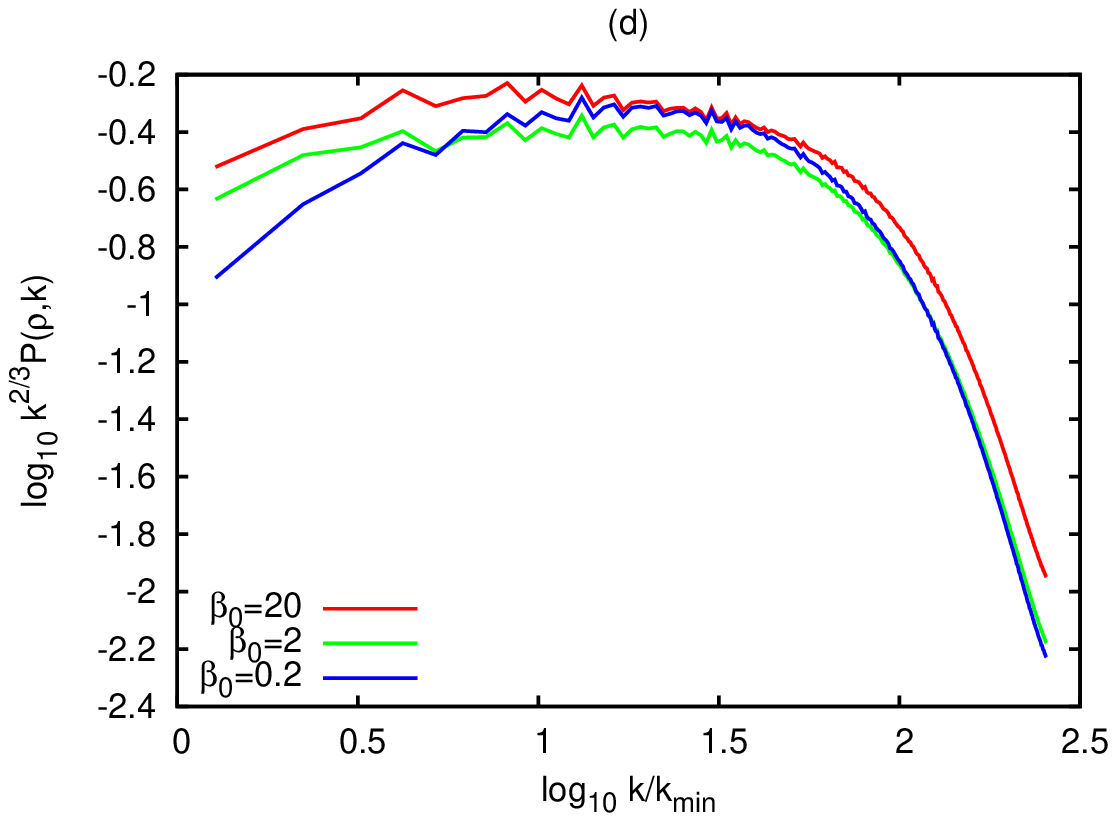}
\plottwo{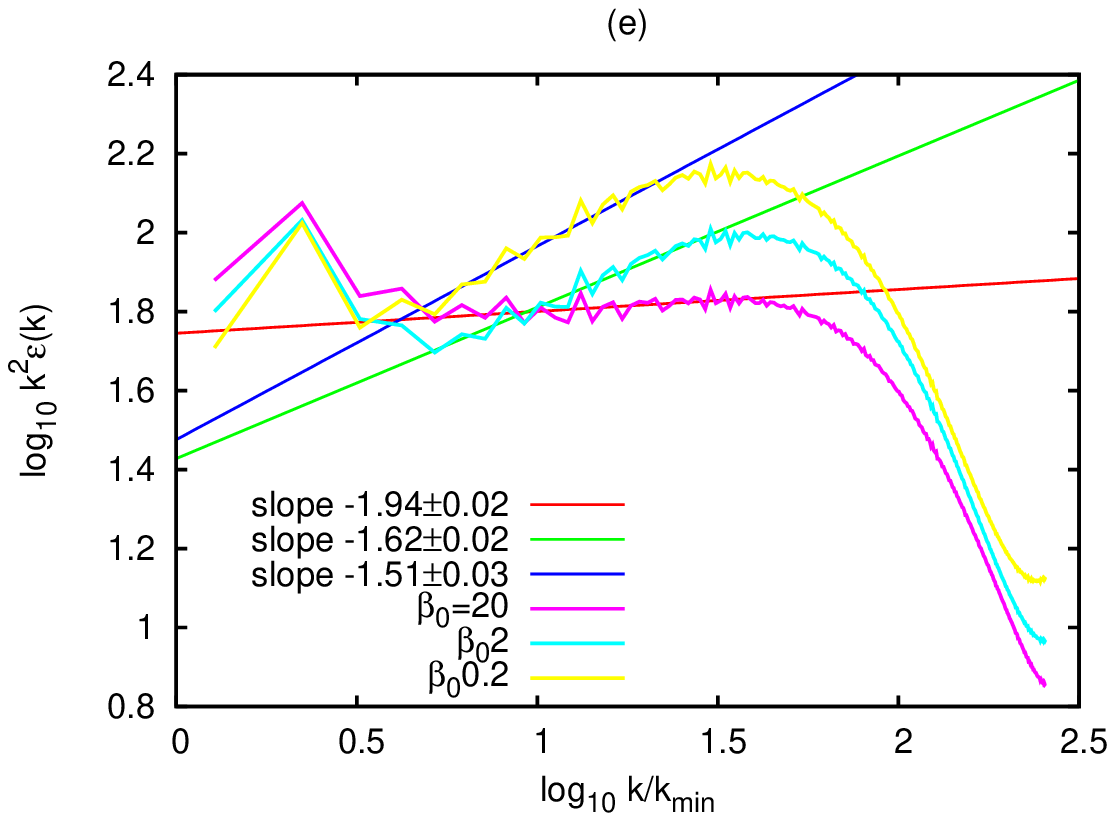}{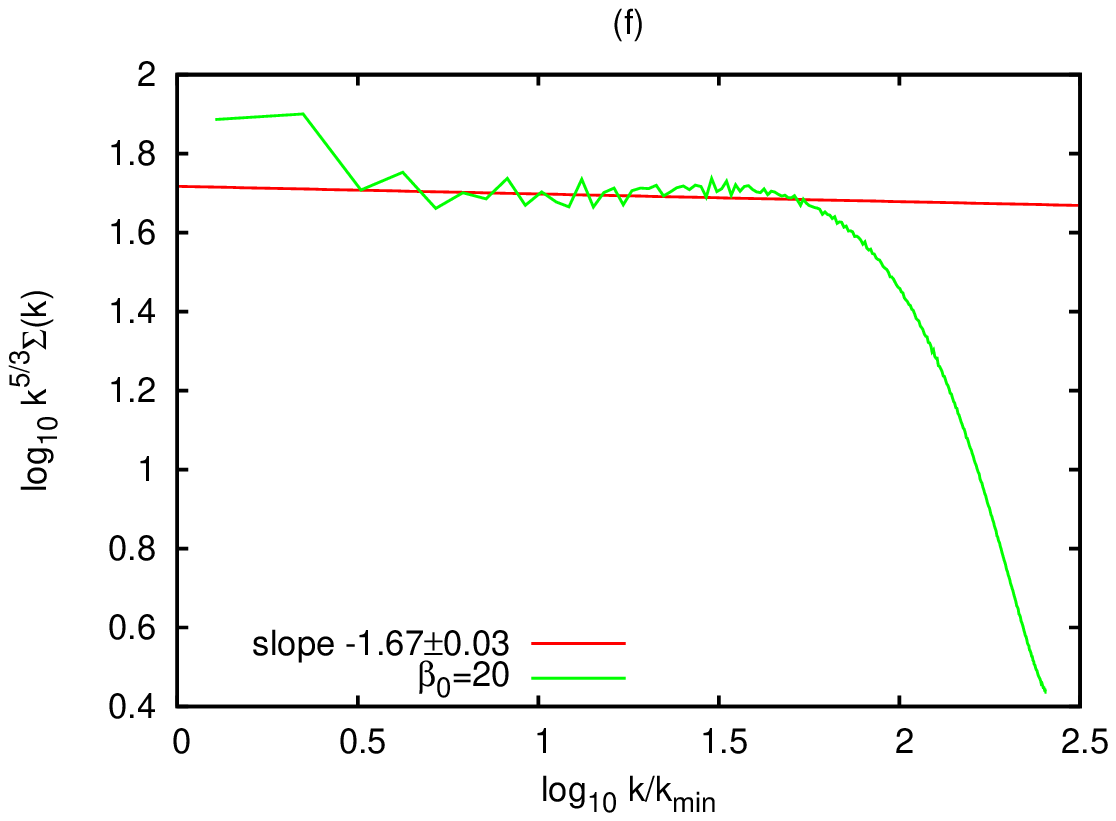}
\plottwo{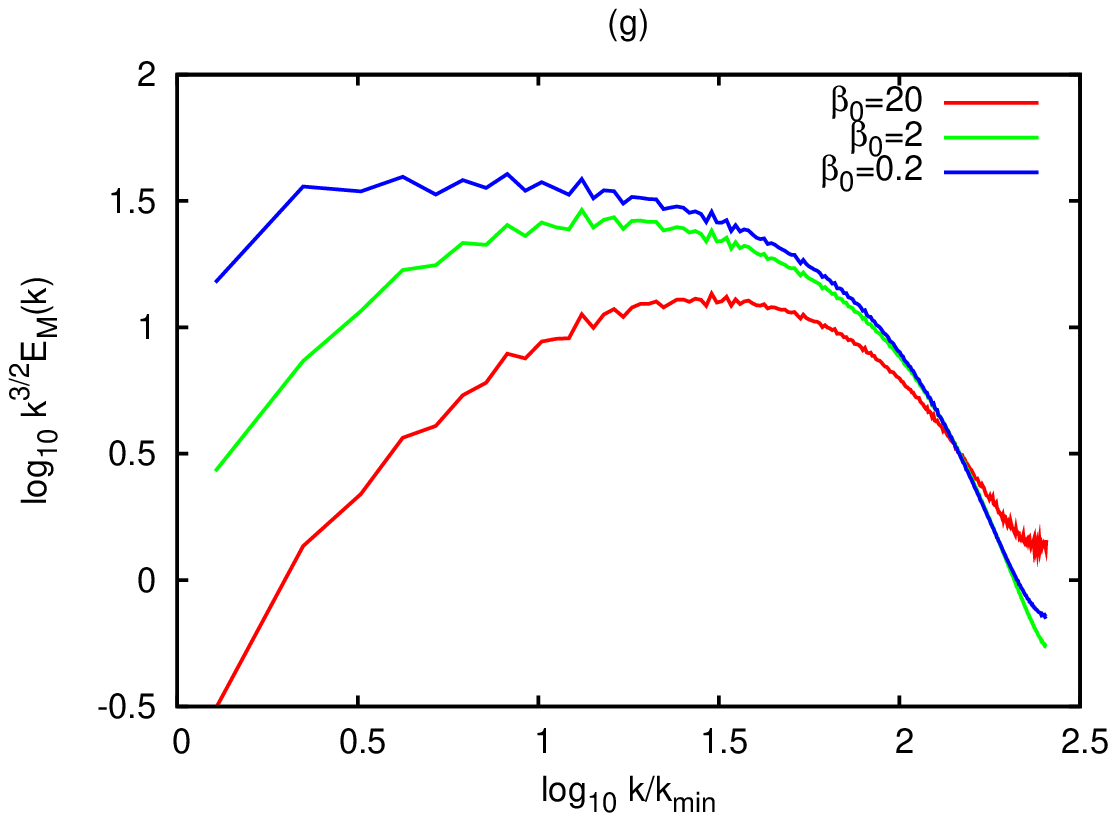}{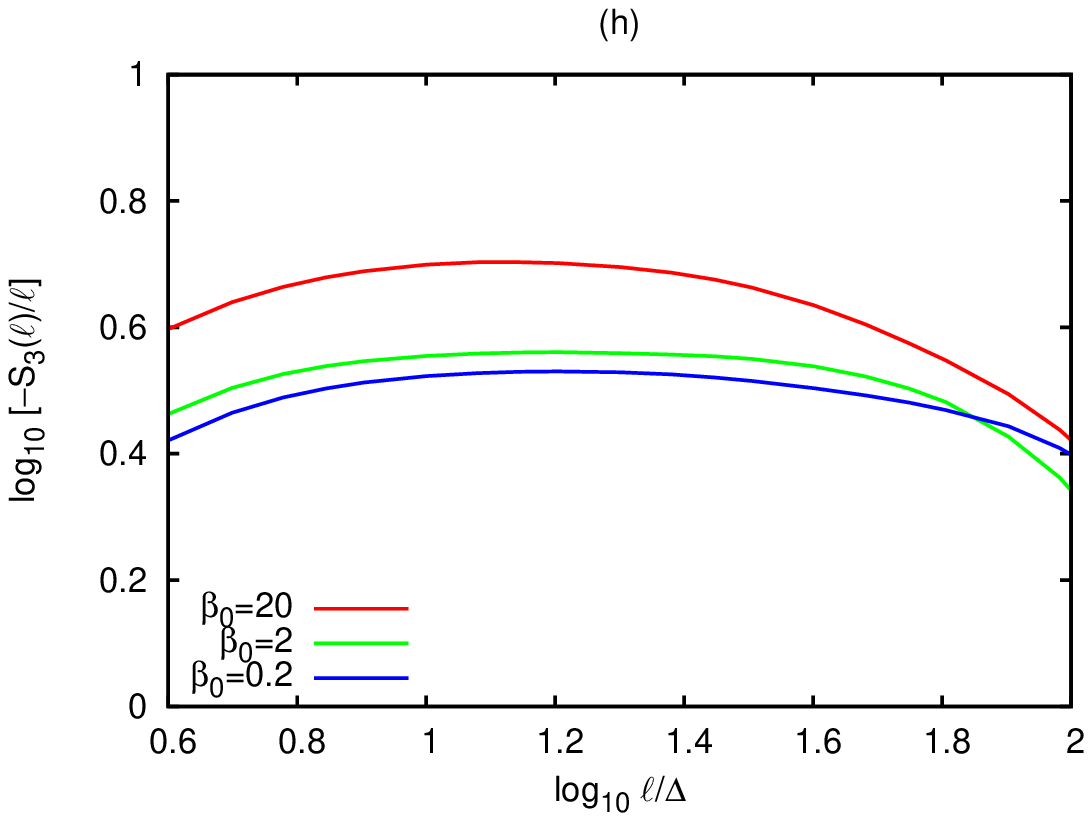}
\caption{\small Statistics of MHD turbulence at $M_s=10$ and $\beta_0=20$, 2, and 0.2 from PPML 
simulations at grid resolution of $512^3$ points: 
({\bf a}) rms sonic and Alfv\'enic Mach numbers vs. time;
({\bf b}) mean kinetic and magnetic energy vs. time;
({\bf c}) minimum gas density vs. time;
({\bf d}) time-average compensated density power spectra;
({\bf e}) time-average compensated velocity power spectra;
({\bf f}) time-average compensated power spectrum of $\rho^{1/3}{\bf u}$ for $\beta_0=20$;
({\bf g}) time-average compensated power spectra of magnetic energy;
({\bf h}) time-average compensated third-order structure functions $S_3(\ell)$ for generalized 
Elsa\"asser fields ${\bf Z}^{\pm}$, see text for the definition details. 
}
\label{figthree}
\end{figure}

We carried out a set of pilot simulations of driven isothermal supersonic turbulence at the
sonic Mach number of 10 on $512^3$ meshes to demonstrate the performance of our new PPML
solver for ideal MHD \citep{popov.07,popov.08,ustyugov...09}. The models are initiated 
with a uniform magnetic
field aligned with the $x$-coordinate direction and cover the transition to turbulence 
and its evolution for up to 10 dynamical times. A large-scale ($k\le2$) nonhelical solenoidal 
force is used to stir the gas in the periodic domain which keeps $M_s$ close to 10 during 
the simulation, see Fig.~\ref{figthree}{\bf a}. The magnetic field is not forced 
and can receive energy
only through interaction with the velocity field. This random forcing does not generate a
mean field, but still leads to amplification of the small-scale magnetic energy 
via a process known as small-scale dynamo \citep[][and references therein]{schekochihin.07}.
It is implicitly assumed that the magnetic Prandtl number $Pm\approx1$ in our numerical models. 
Each simulation reaches a steady state with saturated magnetic energy and a macroscopically
statistically isotropic magnetic field, see Fig.~\ref{figthree}{\bf b}. The properties of this 
saturated state representing fully developed isotropic compressible MHD turbulence is the main 
focus of this section. While some level 
of physical understanding of small-scale dynamo and its saturation exists, the structure of 
the saturated state is poorly known, even in the incompressible case \citep{yousef..07}.

Our models are parameterized by the ratios of thermal-to-magnetic pressure 
$\beta\equiv P_{gas}/P_{mag}$. The initial values $\beta_0=20$ and 2 translate into saturated 
levels of the magnetic energy at about 10\% and 30\% of the kinetic energy, respectively 
(Fig.~\ref{figthree}{\bf b}).
In the $\beta_0=0.2$ case we get a statistical steady state with energy equipartition.
The three simulations fully cover the transition from highly super-Alfv\'enic regime of 
turbulence at $\beta_0=20$ with the root mean square (rms) Alfv\'enic Mach number $M_A\approx10$ through 
mildly super-Alfv\'enic ($\beta_0=2$,  $M_A\approx3$) to trans-Alfv\'enic ($\beta_0=0.2$, 
$M_A\approx 1$), see Fig.~\ref{figthree}{\bf a}. The mean normalized cross-helicity
$\sigma_c\equiv2\left<{\bf u\cdot B}\right>/(\left<{\bf u}^2\right>+\left<{\bf B}^2\right>)$
is contained within $\pm1.5$\% for $\beta_0=20$ and 2, while at $\beta_0=0.2$ the cross-helicity
oscillates somewhat more actively with $\sigma_c\in(-0.073, 0.015)$. Still, to a good approximation, 
the turbulence remains nonhelical.
In all three cases PPML keeps the divergence of the magnetic field at all times to within 
$\left|{\bf\nabla\cdot B}\right|<10^{-12}$. We computed time-average statistics 
over at least four dynamical times for the saturated regimes. The results are discussed below.

The dependence of the density pdf on the value 
of $\beta_0$ shows a very clear trend for substantially weaker rarefactions in the flows 
with stronger magnetic field (smaller plasma $\beta$), see Fig.~\ref{figthree}{\bf c}. 
At a grid resolution of $512^3$, the average minimum density is about $10^{-3.6}$, 
$10^{-2.7}$, and $10^{-2}$ for $\beta_0=20$, 2, and 0.2, respectively. At the same time,
the high-density wing of the pdf preserves its lognormal shape and appears insensitive
to the intensity level of magnetic fluctuations within the studied range of $\beta_0$.

The density power spectra also show some trends with the magnetic field strength, although
they are less pronounced than in the velocity spectra (see below). For example, the power spectrum of 
the logarithm of projected gas density scales as $k^{-1.79}$, $k^{-1.64}$, and $k^{-1.52}$ at 
$\beta_0=20$, 2, and 0.2, respectively (see also Sec.~2 above). The density spectrum for 
$\beta_0=20$  (nearly nonmagnetized medium) at $M_s=10$ scales as $k^{-0.7}$, i.e. it is 
shallower than in our nonmagnetized simulations at $M_s=6$, where the $k^{-1}$ scaling was 
recovered. This result confirms K07 prediction that in the 
limit $M_s\rightarrow\infty$ the density spectrum is flat, $P(\rho,k)\sim k^0$, similar to
the white noise spectrum. At a fixed sonic Mach number
($M_s=10$), as the flow magnetization increases ($M_A$ drops from 10 to 1), the density spectra 
tend to flatten, see~Fig.~\ref{figthree}{\bf d}. The same tendency is clear in the power spectra 
of the logarithm of the density, with the slopes around $-1.3$ at $M_s=10$. This is consistent 
with a slope of $-1.7$ we measured at $M_s=6$ and $\beta_0=\infty$.
We also observe a significant reduction in the extent of the scaling interval in
the $\log{\rho}$ spectra as $\beta_0$ decreases from 20 to 0.2.

We measured a broad range of the velocity power indices from $-1.5$ through $-2$ 
depending on the degree of magnetization (Fig.~\ref{figthree}{\bf e}). As expected, the 
highly super-Alfv\'enic case $\beta_0=20$
is very similar 
to K07 results for nonmagnetized flows, with a Burgers-like scaling of the 
velocity power spectrum, $P({\bf u},k)\sim k^{-1.94}$, and a Kolmogorov-like spectrum 
for the density-weighted velocity, $P(\rho^{1/3}{\bf u},k)\sim k^{-1.7}$ 
\citep[Fig.~\ref{figthree}{\bf f}, also see][]{kowal.07}.
There is a clear trend for the velocity power spectra to get shallower at higher degrees 
of magnetization. We get $k^{-1.62}$ at $\beta_0=2$ and $k^{-1.51}$ at $\beta_0=0.2$, 
consistent with the \citet{lemaster.09} measurement for their strong-field case, $k^{-1.38}$ at 
$\beta_0=0.02$ and $M_s=6.9$. This result suggests that slopes of the velocity power 
spectra around $-1.8$ inferred from the observations of molecular clouds 
\citep[e.g., ][]{padoan...06} may indicate a super-Alfv\'enic nature of the turbulence
there, see however a related discussion in \citet{li...08}.

There is a set of exact scaling laws for homogeneous and isotropic incompressible MHD 
turbulence analogous to the 4/5-law of Kolmogorov for ordinary turbulence in neutral fluids
\citep{chandrasekhar51,politano.98a,politano.98b}. 
The MHD laws can be expressed in terms of Els\"asser fields, 
${\bf z}^{\pm}\equiv{\bf u}\pm{\bf B}/\sqrt{4\pi\rho}$ \citep{elsasser50}, as
$S_{\parallel,3}^{\pm}\equiv\left<\delta z^{\mp}_{\parallel}(\ell) 
[\delta z_i^{\pm}(\ell)]^2\right>=-\frac{4}{d}\epsilon^{\pm}\ell$,
where $\delta{\bf z}_{\parallel}(\ell)\equiv[{\bf z}({\bf x+\hat{e}}\ell)-{\bf z}({\bf x})]\cdot{\bf \hat{e}}$, 
$d$ is the space dimension, ${\bf\hat{e}}$ is a unit vector with arbitrary direction, 
${\bf\hat{e}}\ell$ is the displacement vector, the brackets denote an ensemble average, 
and summation over repeated indices is implied. Equivalently, the scaling laws can be 
rewritten in terms of the basic fields (${\bf u}$, ${\bf B}$), but in MHD there are 
no separate {\em exact} laws for the velocity or the magnetic field alone. It is
important not to neglect the correlations between the ${\bf u}$ and ${\bf B}$ fields
or ${\bf z}^+$ and ${\bf z}^-$ fields constrained by the invariance properties of the 
equations in turbulent cascade models. In supersonic turbulence, it is equally important
to properly account for the density--velocity and density--magnetic field correlations.
In incompressible MHD, the total energy $\left<(u_i^2 + B_i^2)/2\right>$ and the 
cross-helicity $\left<{\bf u}\cdot{\bf B}\right>$ play the role of ideal invariants
and, thus, the (total) energy transfer rate $\epsilon^T=(\epsilon^++\epsilon^-)/2$ 
and the cross-helicity transfer rate $\epsilon^C=(\epsilon^+-\epsilon^-)/2$.
For vanishing magnetic field ${\bf B}$, one recovers the Kolmogorov 4/5-law, {\small
$\left<[\delta {\bf u}_{\parallel}(\ell)]^3\right>=-4/5\epsilon\ell$}, where $\epsilon$ is the mean
rate of kinetic energy transfer \citep{politano.98b}.

Numerical simulations generally confirm these incompressible scalings, although the Rey\-nolds numbers are
perhaps still too small to reproduce the asymptotic linear behavior with a desired precision
and the results are sensitive to statistical errors \citep{biskamp.00,porter..02,boldyrev..06}. 
Often, the absolute value of the longitudinal difference is used and still a linear scaling is
recovered numerically, while there is no rigorous result for the normalization constant in this 
case. The third-order transverse velocity structure functions also show a linear scaling in 
simulations of nonmagnetized turbulence \citep[][K07]{porter..02} and the difference
of scaling exponents for longitudinal and transverse structure functions can serve as a
robust measure of statistical uncertainty of the computed exponents \citep{kritsuk.04}.
In MHD simulations, the third-order structure functions of the Els\"asser fields 
{\small $\left<|\delta z^{\mp}_{\parallel}(\ell)|^3\right>$} were shown
to have an approximately linear scaling too \citep{biskamp.00,momeni.08}, but again there 
is no reason for this result to universally hold in all situations since the correlations 
between the ${\bf z}^{\pm}$ fields play an important role in nonlinear transfer processes.

Can the 4/3-law for incompressible MHD turbulence be extended to highly compressible 
supersonic regimes in molecular clouds? 
As we discussed above, a proper density weighting of the velocity, $\rho^{1/3}{\bf u}$,
preserves the approximately linear scaling of the third-order structure functions at high Mach
numbers in the nonmagnetized case \citep[for a more involved approach to density weighting 
see][]{pan..09}. It is straightforward to redefine the Elsa\"sser fields for compressible
flows using the 1/3-rule, 
{\small ${\bf Z}^{\pm}\equiv\rho^{1/3}({\bf u}\pm{\bf B}/\sqrt{4\pi\rho})$},
so that they match the original ${\bf z}^{\pm}$ fields in the incompressible limit and reduce
to $\rho^{1/3}{\bf u}$ in the limit of vanishing ${\bf B}$. The new ${\bf Z}^{\pm}$ fields
can have universal scaling properties in homogeneous isotropic turbulent flows with a broad 
range of sonic and Alfv\'enic Mach numbers. 

We use data from our MHD simulations to find evidence to support or reject this conjecture 
by computing the third-order structure functions {\small $S_{\parallel,3}^{\pm}$} and {\small 
$S_{\perp,3}^{\pm}$} defined above, but now based on the new {\small ${\bf Z}^{\pm}$} fields. 
Since we are interested in the energy transfer through the inertial interval, we compute the
sum of {\small $S_{\parallel,3}^-$} and {\small $S_{\parallel,3}^+$} which determines the energy transfer rate
$\epsilon^T$ in the incompressible limit. To further reduce statistical errors, we had to 
combine the transverse and longitudinal structure functions, but the
absolute value operator was not applied to the field differences. The results are plotted in 
Fig.~\ref{figthree}{\bf h}, which shows the compensated scaling for {\small $S_3(\ell)\equiv(S_{\parallel,3}^-+
S_{\parallel,3}^+ + S_{\perp,3}^- + S_{\perp,3}^+)/4$} averaged over about two dozen snapshots for
each of the three saturated turbulent states with different levels of magnetic fluctuations.
Although the scaling range is rather short in $512^3$ numerical data, each of the
three cases clearly demonstrates a linear behavior $S_3(\ell)\sim\ell$, in contrast to the 
corresponding scaling of the velocity or the magnetic energy which strongly depend on  
$\beta_0$, see Fig.~\ref{figthree}{\bf e} and {\bf g}. As a consistency check, we computed 
{\small $\tilde{S}_3(\ell)\equiv(S^-_{\parallel,3}+S^+_{\parallel,3})/2$} for both plain ${\bf z}^{\pm}$ fields 
and for density-weighted {\small ${\bf Z}^{\pm}$} fields at $\beta_0=2$. We then compared the scaling 
$\tilde{S}_3\sim\ell^{\,\alpha}$ for the two cases and found that the plain Els\"asser fields result
in a steeper slope, $\Delta\alpha\equiv\alpha_{\bf z}-\alpha_{\bf Z}\approx0.23$. K07 measured
a similar slope difference, $\Delta\alpha\approx0.31$, for the third-order longitudinal 
velocity structure functions in nonmagnetized turbulence at $M_s=6$.

While larger simulations are definitely needed to confirm the linear scaling, we believe that our
preliminary evidence suggests an interesting new approach to tackle universal scaling relations
and intermittency in compressible MHD turbulence.

\section{Conclusions}

We further supported the 1/3-rule of K07 for hydrodynamic supersonic turbulence 
with new data from a larger simulation at $M_s=6$ on a grid of $2048^3$ points. 
Based on a series of $512^3$ MHD turbulence simulations at $M_s=10$, we explored 
universal trends in scaling properties 
of various statistics as a function of the magnetic field strength. Our data 
suggest that the 4/3-law of incompressible MHD can be extended to supersonic flows.

\acknowledgements {\small
This research was supported in part by the National Science Foundation through 
grants AST-0607675 and AST-0808184, as well as through TeraGrid resources 
provided by NICS, PSC, SDSC, and TACC (MCA07S014 and MCA98N020).}

%%% THE BIBLIOGRAPHY

\end{document}